# Weak Localization and Dimensional Crossover in Carbon Nanotube Systems

M. Salvato[a], M. Lucci, I. Ottaviani, M. Cirillo

*Dipartimento di Fisica and MINAS Laboratory, Università di Roma "Tor Vergata", I- 00133 Roma*

[a] Also *CNR-SPIN* Institute, Italy

S. Orlanducci, F. Toschi, and M. L. Terranova

*Dipartimento di Scienze e Tecnologie Chimiche and MINAS Laboratory, Università di Roma "Tor Vergata", I- 00133 Roma*

## Abstract

We investigate the effects of magnetic and electric fields on electron wavefunction interactions in single walled carbon nanotube bundles. The magnetoresistance measurements performed at *4.2K* and the dependence of the data upon the electric field, obtained by varying the bias current through the samples, reveal good agreement with weak localization theory. Recording current-voltage characteristics at different temperatures we find an ohmic non-ohmic transition which disappears above *85K*. Conductance vs temperature measurements are also well explained in the framework of weak localization theory by the predicted temperature dependence of the electric field-conditioned characteristic length. This length results equal to the average bundles diameter just at *T≅85K*, indicating that the observed conductance transition is due to a 2D-3D crossover.

Much interest is devoted today to the study of mesoscopic effects, namely phenomena that can be classified as macroscopic for the nature of their observation, but are explained as evidence of the microscopic (quantum) structure of the systems under consideration. Books [1] and reviews [2] have appeared on this topic which is still growing and offering new and interesting developments both at fundamental and applied physics level. Within this framework the study of electron transport phenomena in low-dimensional conductors, like carbon nanotube compounds [3] and new superconducting materials and devices [4], has achieved a relevant role. Among the theories and the models conceived to give account for experimental data in terms of electron wavefunctions properties and interactions in low-dimensional systems the Weak Localization theory (WL theory) [5] occupies a role which has been relevant and productive.

WL theory, based on the assumption that the electron wavefunctions extend over a length determined by the disorder inside the conductors, has been successfully used for explaining the properties of disordered metallic [6] and semiconducting films [7] as well as aggregates of carbon nanotube bundles [8,9]. The theory assumes that charge carriers are localized on a diffusion length $L=(D\tau_i)^{1/2}>l_0$, where $D$ is the electrons diffusion constant, $\tau_i$ the inelastic mean free time and $l_0$ the elastic mean free path. External fields and temperature effects contribute to delocalize the electron wavefunctions affecting the diffusion length $L$ and causing a change in the sample conductivity[10]. WL theory also predicts that the sample dimension $d$ plays a role in determining the electrical conductivity: depending on whether the value of the ratio $d/L$ is lower or greater than the unity, respectively *2* or *3* dimensional (2D or 3D) effects can be evidenced. In this paper we intend to demonstrate that, for a system of Single Walled Carbon Nanotubes (SWCNT), the theory of weak localization can provide a satisfactory explanation of the effects generated by external magnetic and electric fields; moreover we will show that, tuning the diffusion length by the electric field generated by the currents, a 2D-3D crossover can be observed when this length matches the physical dimensions of the SWCNT bundles around the temperature of *85K*.

Aggregates of SWCNT bundles have been deposited in our laboratory on Si-SiO substrates by dielectrophoresis techniques [11]. On the substrates, prior to the deposition, we patterned four gold contacts acting as current and voltages electrodes as shown in the Scanning Electron Microscope (SEM) image of Fig. 1a. The distance between the two voltage probes (the internal ones) is $l_V=5\mu m$ while the distance between the two current electrodes is $l_c=60\mu m$. The average diameter of each SWCNT is $d_{SWCNT}=1.2nm$ and the diameter of the bundles, measured by SEM is $d_B=80\pm20nm$. The SWCNT bundles had lengths ranging in the interval $(1-10)\mu m$ and the SEM analysis evidenced that the bundles were indeed deposited in the form of aggregates bridging the contacts pads; some snapshots of this analysis are shown in Fig. 1b. In Fig. 1b we can clearly see the bundles as the white filaments and from the picture we also have an idea of the three dimensional nature of the aggregate. Clearly visible in the figure are also the joints where the bundles overlap and are in contact because in these points we obtain higher intensity of the signal. The measurements herein reported are very typical of a set of a dozen samples that we investigated

The temperature measurements in the range (*4K-300K*) were performed by thermally anchoring the samples to the cold finger of a high vacuum cryocooler while the Magneto Resistance (MR) measurements have been performed keeping the samples in He liquid bath (*T=4.2K*) inside a cryostat equipped with a *6T* superconducting magnet. Fig.2a shows the changes of the MR of a sample scaled by the zero-field resistance value and normalized to this same reference value: we plot *(R(B,T)-R(0,T))/R(0,T)*, where *R(B,T)* and *R(0,T)* are the sample resistances measured with and without the external magnetic field respectively, at *T=4.2K* obtained with different dc bias currents. The negative MR observed at low field is indicative of 2D weak localization regime[8] and is given by

$$\frac{\Delta R}{R(0,T)} = -R(0,T)\frac{e^2}{2\pi\hbar}\left\{\psi\left(\frac{1}{2}+\frac{\hbar}{4eL_H^2 B}\right) - \ln\left(\frac{\hbar}{4eL_H^2 B}\right)\right\} \quad (1)$$

where *ΔR=R(B,T)-R(0,T)*, *e* is the electron charge, $\hbar$ is the Planck constant over *2π*, *Ψ* the di-Gamma function, *B* the external magnetic field and $L_H=(D\tau_H)^{1/2}$ is the diffusion length associated to the magnetic delocalization and used as fit parameter; the parameter $\tau_H$ is the inelastic mean free time due to the external magnetic field. The data fitting (continuous curves in Fig. 2a) gives a constant value for $L_H=30nm$ at low bias current with a decrease when the bias current increases above *10 µA* ; The typical error bar for the vertical component of the data in Fig. 2a range between of one fifth (±0.001) and one half (±0.0025) of the small ticks spacing. These errors are not relevant for the fittings indeed because we see that the theoretical curves are right on the top of the experimental data. In Fig. 2b we report, for each value of the bias current, the values of $L_H$ used for the fit and the corresponding value of $\tau_H$ , obtained assuming $D=50cm^2/s$ [8]. Its value results higher than the elastic scattering time $\tau_0=10^{-15}s$ estimated by literature data [8] in agreement with the condition for WL which requires $\tau_H > \tau_0$. At *T=4.2K* the condition for MR upturn, given by $g\mu_B B/k_B T >> 1$ [10] (with $\mu_B$ the Bohr magneton, *g=2* the gyromagnetic ratio and $k_B$ the Boltzmann constant), is $B \cong 3T$.

The above calculated value of the magnetic induction is consistent with the experimental result shown in Fig. 2a. The upturn of the MR disappears for high bias currents indicating that the external driving force provided by the electric fields associated with the bias current affects the charge localization. We note that the effect of the bias current (or electric field) generates the same kind of features observed increasing the temperature of the samples [6] (the temperature is fixed at *4.2K* for the measurements of Fig. 2a). In other word an increasing electric field can delocalize the electrons just as the temperature can do it ; this phenomenon had been predicted [8], however, evidence of it had not been reported so far. It is worth noting that the MR measurements revealed negligible anisotropy of the bundles aggregates (this peculiarity can be expected from the pictures of Fig. 1b). Indeed we recorded very little differences in MR measurements when the external field was directed parallel and perpendicular to the electrodes shown in Fig. 1a: for space limitation these specific measurements and comparison are not be reported herein.

The effect of an electric field on the localization of electrons was investigated biasing the samples, in zero magnetic field, with a dc current and recording the current-voltage (*I-V*) characteristics at different temperatures; the experimental results obtained in this case are shown in Fig.3a. Increasing the current, a transition between an ohmic (*I*~*V*) to a non-ohmic (*I*~*V²*) regime is clearly observed at low temperature with a crossover voltage $V^*$ : we show this effect in Fig. 3a for the current-voltage curve obtained at 4K where we also show how $V^*$ is determined. The crossover voltage increases with the temperature. Increasing the temperature, the non-ohmic regime is confined at higher current and disappears when *T>85K*. A diffusion length is associated to the external electric field in weak localization regime and is given by[12]:

$$L_F = \left(\frac{\hbar D}{eF}\right)^{1/3} \qquad (2)$$

where *F* is the electric field between the current electrodes. The substitution of $F^*=V^*/l_c$ inside this expression allows to obtain the temperature dependence of $L^*_F$ which gives the lower limit of the diffusion length for ohmic behaviour. This dependence is reported in Fig. 3b where the value $L^*_F=93nm$ is obtained for *T=85K*. This value is of the same order of the bundle diameter $d_B$ suggesting that the ohmic-non ohmic transition is strictly connected to the dephasing scattering of the charge carriers wavefunction with the bundle walls when $L_F \cong d_B$. The condition $L_F \lesssim d_B$ also characterizes the 2D-3D WL crossover with $T \cong 85K$ the crossover temperature in agreement with data reported by other authors [8,9,13]. In Fig. 3b we also show that the $L_F$ dependence upon the temperature follows quite closely a *1/T* dependence which is represented by the continuous curve in the figure. We shall see later that this functional behaviour is consistent with the experimental observations (and fittings) related to the current-voltage characteristics and to the 2D-3D crossover.

In order to better characterize the temperature and current dependence of this ohmic-non ohmic crossover, conductance (*G=1/R*) vs. *T* measurements at different bias currents and zero applied magnetic field have been performed. Fig. 4a and 4b show the experimental data obtained at low current ($I=0.5\mu A$) and at high current ($I=100\mu A$) respectively. In both the figures experimental data have been fitted with the theoretical curves obtained for 2D and 3D weak localization theory [5,14]:

$$G_{2D}(T) = G_0 + \frac{e^2}{2\pi^2 \hbar} \frac{S}{l} \ln\left[1 + \left(\frac{T}{T_0}\right)^p\right] \quad (3a)$$

$$G_{3D}(T) = G_0 + \frac{e^2}{\pi^3 \hbar} \frac{S}{l} \frac{1}{a} T^{p/2} \quad (3b)$$

where $G_0$ is the zero temperature limit conductance obtained assuming only elastic scattering by impurities, $S$ and $l$ are the surface section and the length of the sample respectively and a temperature dependence of the diffusion length $L_{Th}=aT^{p/2}$ is assumed.

The value $I=0.5\mu A$ polarizes the sample in the ohmic-non ohmic transition region observed below $85K$ in the I-V characteristics of Fig.3a. Increasing the temperature above $85K$ the ohmic-non ohmic transition region is crossed. Biasing the sample with $I=100\mu A$, the ohmic region is explored without any crossing between the two regions. As a result, at low bias the data of Fig.4a are well fitted by the 2D WL curve for temperature below $85K$ and with the 3D WL curve for $T>85K$. At high bias current, on the other hand, the 3D WL curve represents the best fit of the data over the whole temperature range ($4K-250K$). These results are in agreement with the temperature dependence of $L_F$ shown in Fig.3b which suggests a 3D regime at high bias when $L_F<d_B$ and a 2D regime at low bias when $L_F>d_B$. The 2D-3D crossover observed in Fig.4a is due to the increasing temperature in an otherwise 2D regime. At high bias, the already established 3D WL regime does not allow any temperature dependence crossover. The exponent $p$ for the 2D model, obtained by the fitting procedure, is $p=2.12$ which indicates an electron-electron scattering process in the non-ohmic region [10,15,16]. The electron-electron scattering is favoured by the lowering of the dimensionality which increases the interaction probability between the carriers [5,10]. We also note now that the two above expressions (3a) and (3b) are derived under the assumption that the average lifetime for inelastic scattering depends on the temperature according to the relation $\tau_i \propto T^{-p}$.

Thus, using p as a fitting parameter for the G vs T data it is possible to obtain information on inelastic scattering. We have seen that the data of Fig. 4a return $p \cong 2$ which is the exponent expected for electron-electron scattering [16]; the diffusion length $L_F$ is related to the time by $L_F = \sqrt{D\tau_i} \propto T^{-p/2}$ which, for $p=2$ gives $L_F \propto T^{-1}$ which is the behaviour that we have obtained and shown in Fig. 3b.

Our data show that the dimensional crossover is governed by $L_F$ and takes place when $L_F \cong d_B$ where $d_B$ is the bundle diameter which is larger than the SWCNT diameter $d_{SWCNT}$. This suggests that most of the dissipative processes are due to the bundles-bundles interface in agreement with the hypothesis that at the interfaces the highest potential barriers are localized. This result confirms our previous observations and experimental results[17] on the importance of the role played by bundles-bundles interfaces on the SWCNT conductivity with respect to the interfaces contacts of the inner SWCNT. The SWCNT-SWCNT interfaces inside each bundle, on the other side, present a much lower energy barrier which can be crossed by carriers with minor effect on transport properties [18].

In conclusion we have demonstrated that the theory of weak localization can provide a very complete picture of the transport phenomena in carbon nanotubes aggregates. We have shown that the localization length is indeed a function of magnetic field, electric field and temperature and we have illustrated how a 2D-3D WL crossover can be induced in SWCNT bundles aggregates by temperature and current bias variation. External fields and temperature effects contribute to delocalize the electron wavefunctions affecting the diffusion length *L* and causing a change in the sample conductivity. Consideration of the peculiar structural characteristics has allowed us to confirm previous models on the origin of the dissipation in the aggregates of SWCNT.

This work has been partially supported by the *GESTO* program of the Regione Lazio, Italy.

References

[1] Y. Imry, *Introduction to Mesoscopic Physics*, second edition (Oxford Univ. Press, New York,2002); E. Akkermans and G. Montambaux, *Mesoscopic Physics of Electrons and Photons* (Cambridge Univ. Press, New York 2007).

[2] F. A. Buot, Physics Report **234**, 73 (1993); Y. Imry and R. Landauer, Reviews of Modern Physics **71**, S306 (1999); P. Sheng and B. van Tiggelen, Waves in Random and Complex Media, Volume **17**, 235 (2007).

[3] A.B. Kaiser Rep. Prog. Phys **64**, 1 (2001);Y. Yosida, I. Oguro Appl. Phys. Lett. **86**, 999(1999).

[4] M. Ferrier, A. Kasumov, R. Deblock, S. Guéron, and H. Bouchiat, Journal of Physics **D**: Applied Physics (IOP, October 2010).

[5] P.A. Lee, T.V. Ramakrishnan, Rev. Mod. Phys. **57**, 287(1985).

[6] D.E. Beutler, N. Giordano, Phys. Rev. **B38**, 8(1988).

[7] V. Bayot, L. Piraux, J.P. Michenaud, J.P. Issi, Phys. Rev. **B40**, 3514(1989).

[8] S.N. Song, X.K. Wang, R.P.H. Chang, J.B. Ketterson, Phys. Rev. Lett. **72**, 697(1994).

[9] M.S. Fuhrer, W. Holmes, P.L. Richards, P. Delaney, S.G. Louie, A. Zettl, Synthetic Metals **103**, 2529(1999).

[10] G. Bergmann, Physics Report **107**, 1(1984).

[11] M.L. Terranova et al. , J.Phys.D: Condensed Matter **19**, 2255004 (2007).

[12] M. Kaveh, M.J. Uren, R. A. Davies, M. Pepper J. Phys. C: Solid State Phys. **14**, 413(1981).

[13] G. Baumgartner, M. Carrard, L. Zuppiroli, W. Bacsa, W.A. Heer, L. Forrò, Phys. Rev. **B55**, 6704(1997); M. Baxendale, V.Z. Mordkovich, S. Yoshimura, Phys. Rev. **B56**, 2161(1997); G.T. Kim, E.S. Choi, D.C. Kim, D.S. Suh, Y.W. Park, Phys. Rev. **B58**, 16064(1998).

[14] L. Langer, V. Bayot, E. Grivei, J.-P. Issi, J.P. Heremans, C.H. Olk, L. Stockman, C. Van Haesendonck, Y. Bruynseraede, Phys. Rev. Lett. **76**, 479(1996).

[15] P. K. Choudhury, M. Jaiswal, R. Menon, Phys. Rev. **B76**, 235432(2007).

[16] N. W. Ashcroft and N. D. Mermin, *Solid State Physics* (Holt Rinehart & Winston, New York, 1976).

[17] M. Salvato, M. Cirillo, M. Lucci, S. Orlanducci, I. Ottaviani, M.L. Terranova, F. Toschi Phys. Rev. Lett. **101**, 246804 (2008).

[18] H. Xie, P. Sheng, Phys. Rev. **B79**, 165419 (2009).

**Figure Captions**

Fig. 1 SEM images of our samples (a) the four gold probe configuration and (b) details of our single-walled carbon nanotube (SWCNT) bundles aggregates contacting the gold electrodes: each bundle is represented by a single white filament in the photos. We can see from the scales that the dimensions of the bundles are of the order of 100nm ; the photos also give the idea that two or three superimposed level of aggregates are present. Also well visible are the junctions between the bundles.

Fig.2 (a) Magnetoresistance vs. external magnetic field for different bias current. The lines are fits to the data using equation (1) ; (b) the dependence of the fitting parameters for Eq. 1 as a function of the bias current: the empty circles are relative to the inelastic scattering time $\tau_H$ values while the corresponding $L_H$ values are indicated by the full squares.

Fig.3 (a) *I-V* characteristics at different temperatures. The intersection between the straight lines is representative of the method used for the determination of $V^*=F^*l_c$ in equation (2); (b) temperature dependence of the diffusion length $L_F$ obtained from equation (2). The horizontal dashed line is at the value of $L_F$ when *T=85K* and represents the 2D-3D crossover boundary.

Fig.4 (a) Conductance vs. temperature measurements in the case of a) *I=0.5μA* and b) *I=100μA*. The lines are fit to the data using equation 3a (dashed line) and 3b (continuous line) in the range (*4-85) K* and (*85-300)K* respectively; (b) the fact that the data are fitted in the whole temperature range according to the equation (3b) confirms the 3D WL behaviour for the higher current bias.

**(a)**

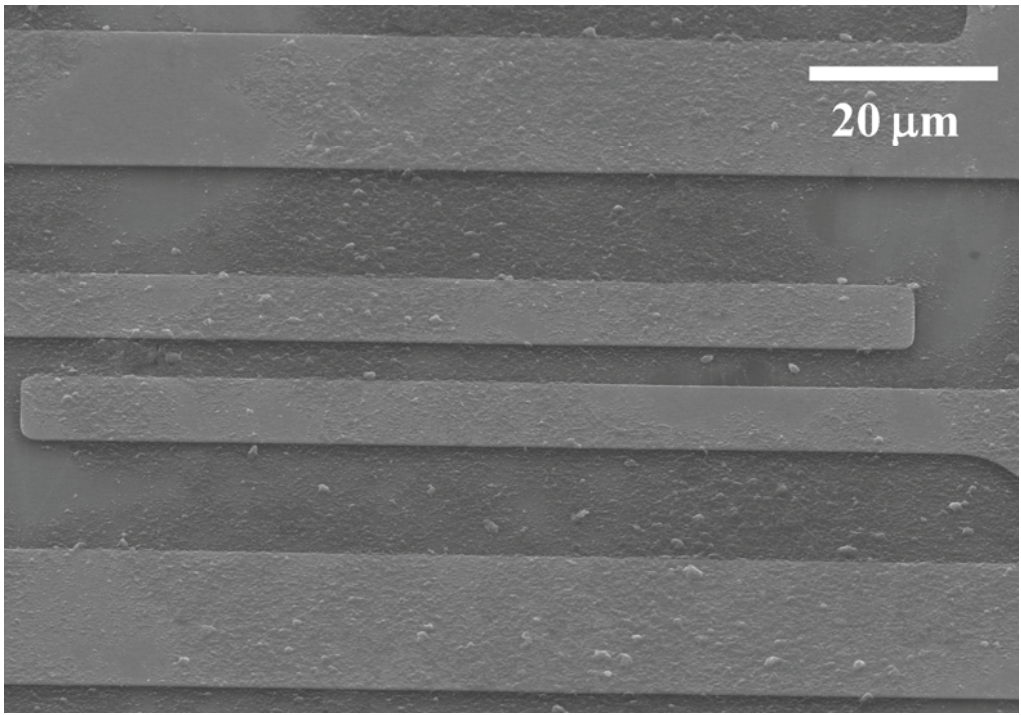

**(b)**

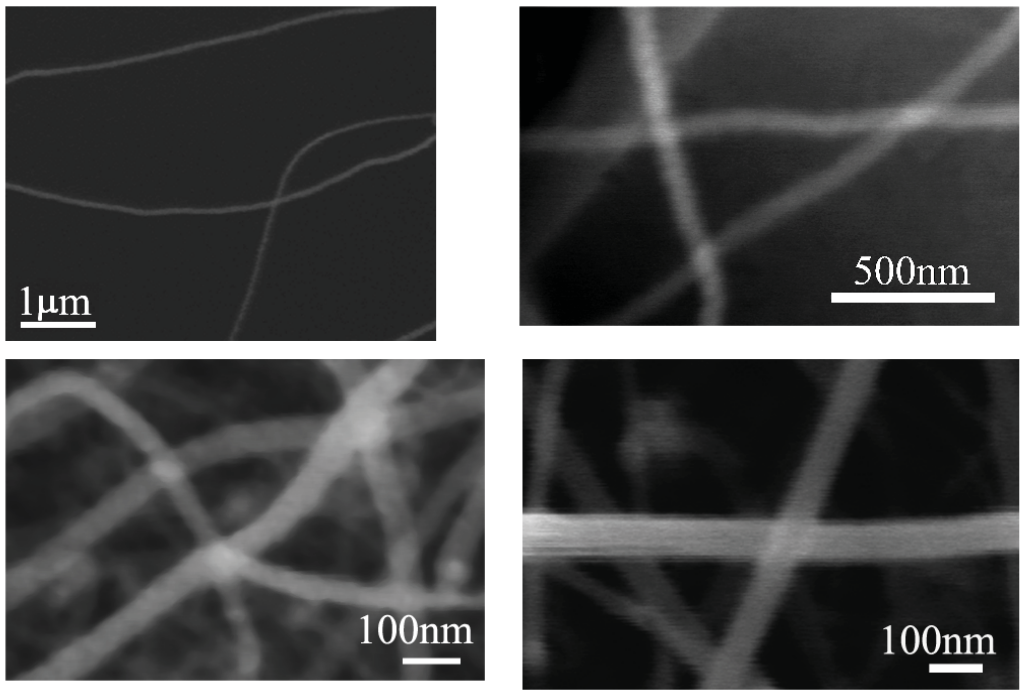

Figure 1, M. Salvato et al.

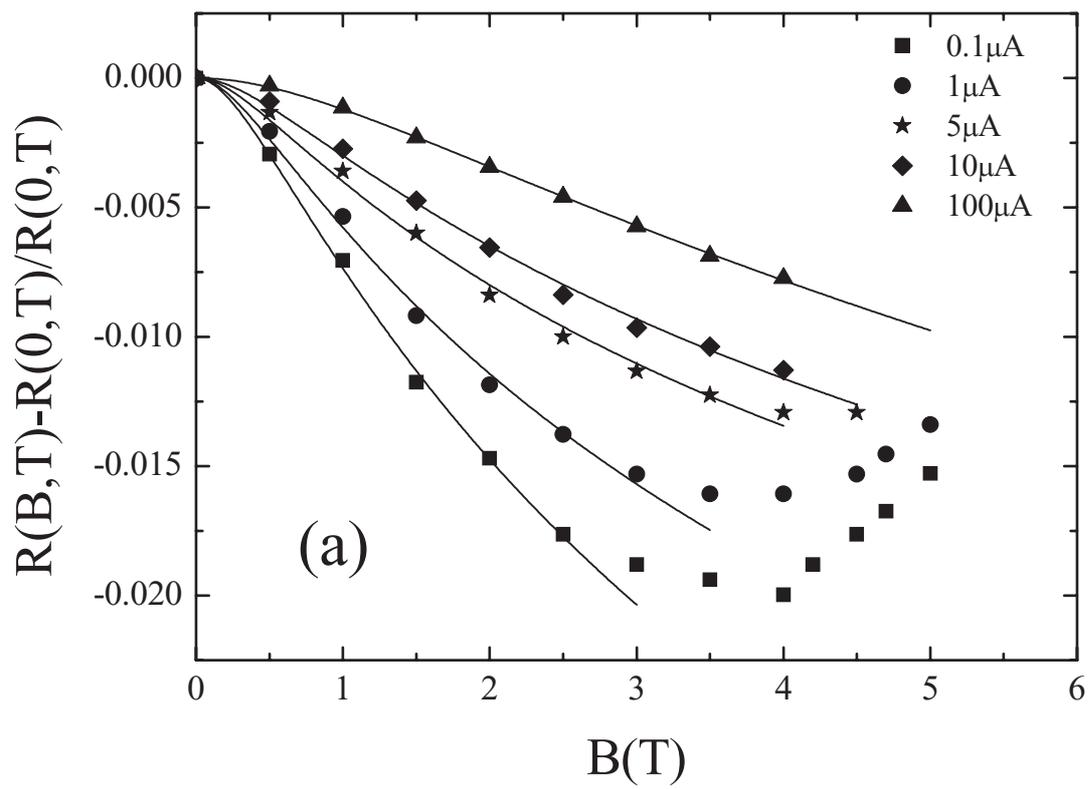

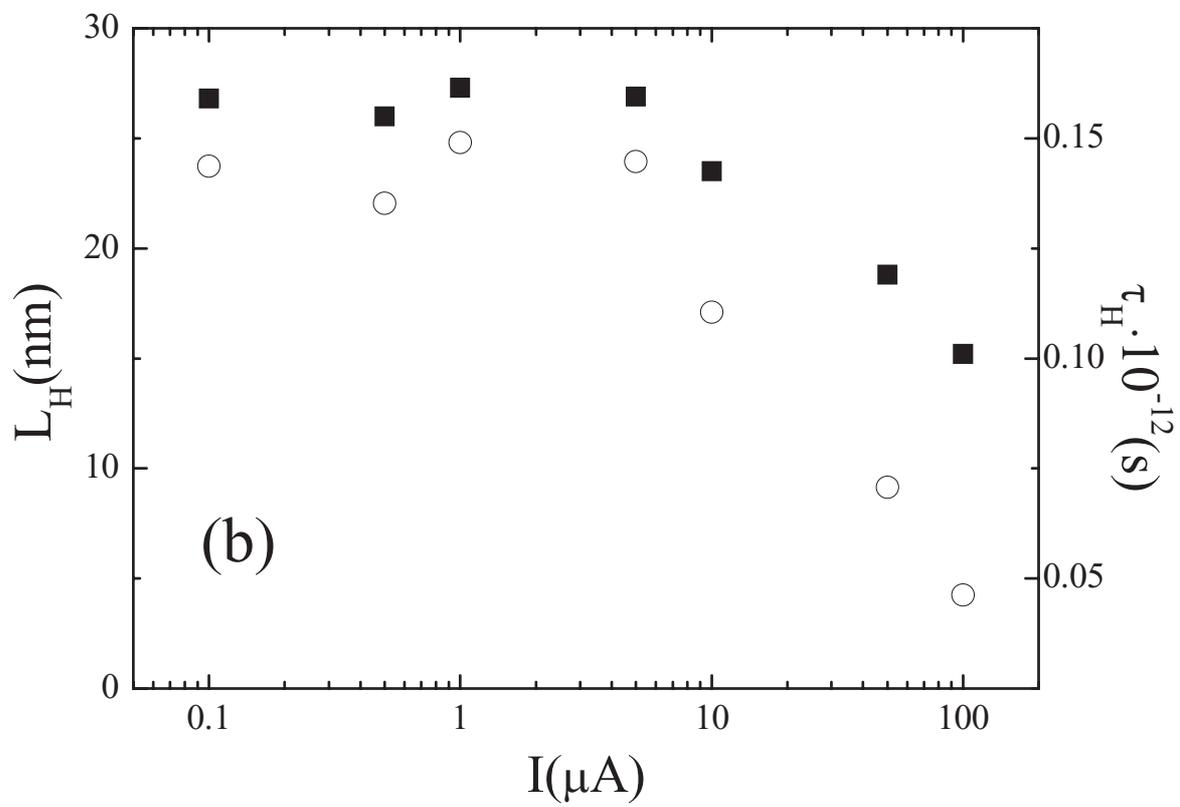

Figure 2, M. Salvato et al.

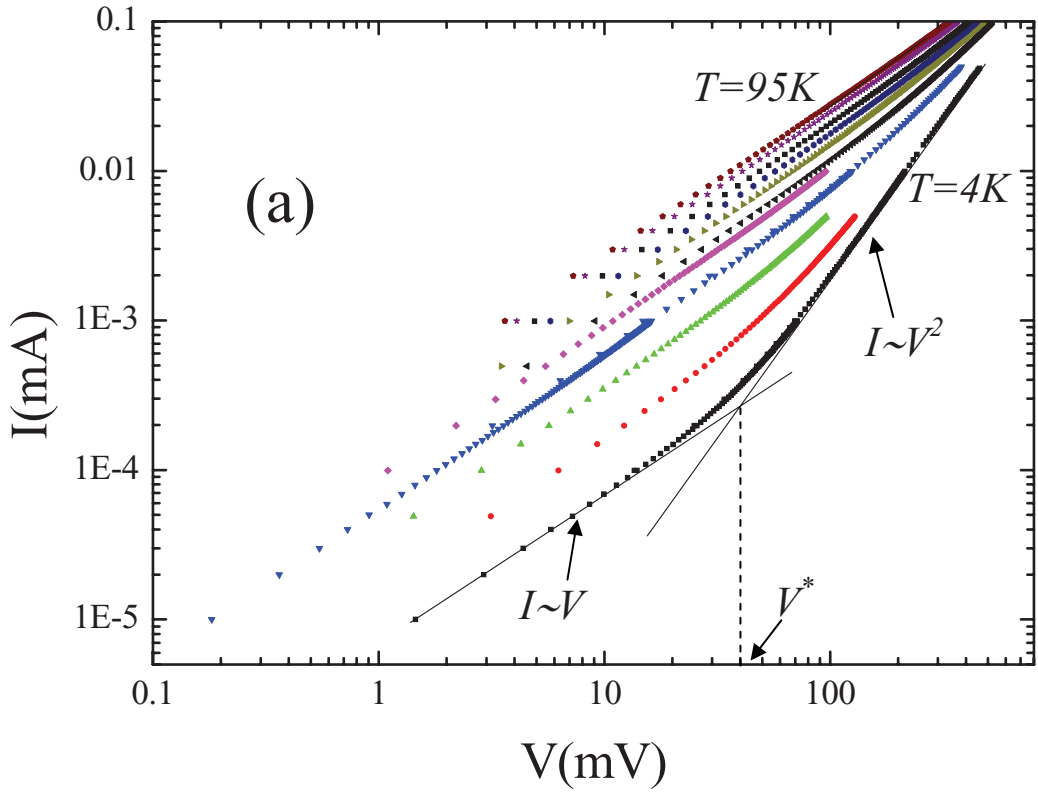

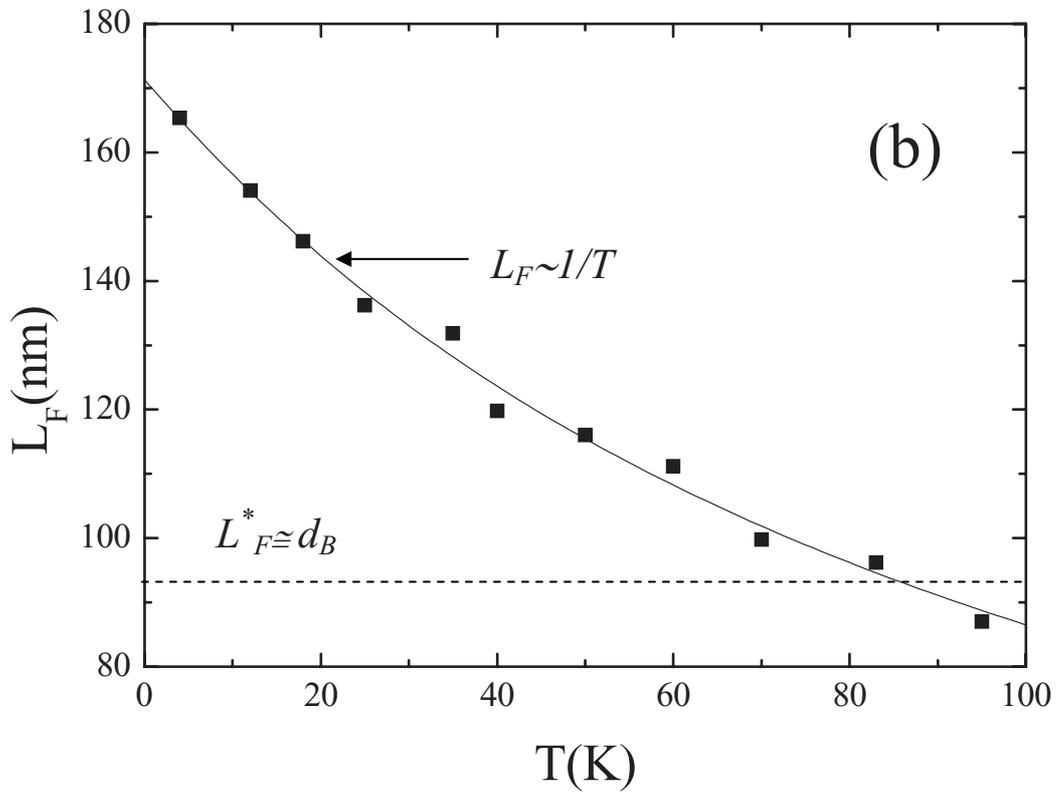

Figure 3, M. Salvato et al.

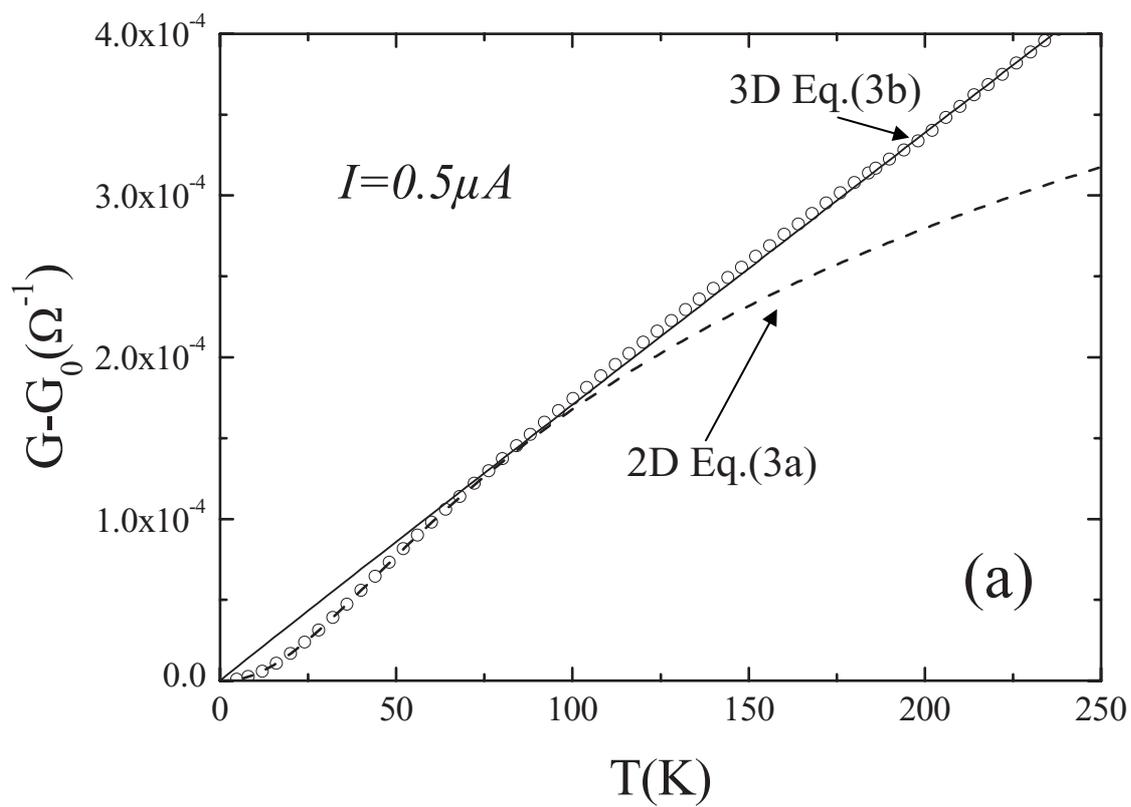

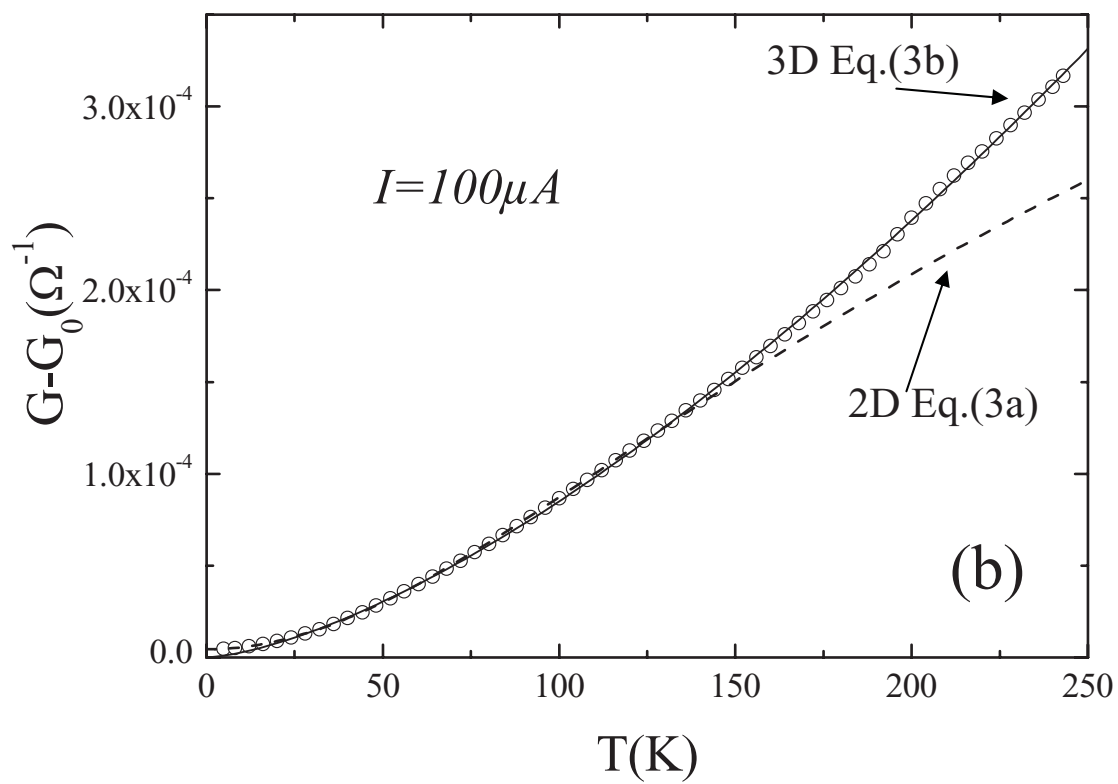

Figure 4, M. Salvato et al.